# End-to-end Evaluation of Practical Video Analytics Systems for Face Detection and Recognition

*Praneet Singh, Edward J Delp, Amy R Reibman*
*Elmore School of Electrical Engineering, Purdue University*

## Abstract

*Practical video analytics systems that are deployed in bandwidth constrained environments like autonomous vehicles perform computer vision tasks such as face detection and recognition. In an end-to-end face analytics system, inputs are first compressed using popular video codecs like HEVC and then passed onto modules that perform face detection, alignment, and recognition sequentially. Typically, the modules of these systems are evaluated independently using task-specific imbalanced datasets that can misconstrue performance estimates. In this paper, we perform a thorough end-to-end evaluation of a face analytics system using a driving-specific dataset, which enables meaningful interpretations. We demonstrate how independent task evaluations, dataset imbalances, and inconsistent annotations can lead to incorrect system performance estimates. We propose strategies to create balanced evaluation subsets of our dataset and to make its annotations consistent across multiple analytics tasks and scenarios. We then evaluate the end-to-end system performance sequentially to account for task interdependencies. Our experiments show that our approach provides consistent, accurate and interpretable estimates of the system's performance which is critical for real-world applications.*

## Introduction

End-to-end face analytics systems deployed in the real-world sequentially perform multiple tasks like compression, detection, alignment, and recognition. In practical systems, these modules operate interdependently in order to meet resource and bandwidth constraints. The performance of a given module in such systems usually depends on how well the previous modules performed. For example, for practical face analytics, the performance of the recognition module will always depend on how well faces are detected by the detection module.

Until now, each component of a practical end-to-end analytics system has been evaluated independently without considering this task interdependence. Furthermore, the datasets used for evaluation of such systems are task-specific, i.e., they lack annotations that are suitable for evaluating multiple tasks like face detection and recognition together. Also, these datasets usually consist of images scraped from the web and they are imbalanced in terms of characteristics like samples per identity, face poses, and lighting conditions. For these reasons, it is difficult to obtain meaningful and realistic estimates of the system's performance.

In addition, practical video analytics systems rely on complex deep learning models to perform several computer vision tasks. However, due to resource and bandwidth constraints that exist in practical scenarios, it is necessary to evaluate: 1) the efficacy of lightweight models in practical analytics, and 2) the effects of compression on each module of an end-to-end system.

In this paper, we perform an end-to-end evaluation of a practical video analytics system to obtain meaningful, interpretable results. To do so, we design a face analytics system (Figure 1) and evaluate its performance on a driving-specific face dataset. This dataset captures vehicle occupants in real-world scenarios with different camera views, modalities, and in different illumination conditions. During evaluation, we show that it is necessary to consider the following so that an accurate estimate of overall system performance can be obtained:

1. The performance interdependence between analytics tasks;
2. The dataset is balanced across all evaluation scenarios;
3. The dataset annotations are consistent for evaluating multiple face analytics tasks sequentially.

Here, we first demonstrate how failing to adhere to the above cases can produce false performance estimates that can be highly misleading. Then we propose strategies that help overcome these challenges and demonstrate that to obtain meaningful and interpretable performance estimates for practical video analytics systems requires (a) end-to-end assessment (b) using balanced datasets (c) with consistent annotations that (d) incorporates the impact of resource and bandwidth constraints.

## Background

For face detection, learning-based algorithms proposed in [1, 2, 3] are extremely effective. These algorithms use simple similarity transforms to align the faces using facial landmarks. These algorithms are evaluated on popular datasets like WiderFace [4] and FDDB [5] that are imbalanced and solely annotated for face detection. Hence, they cannot be used for the evaluation of face recognition systems.

For face recognition, approaches proposed in [6, 7, 8] help learning-based models achieve the best performance. In the case of recognition, the models are evaluated on perfectly aligned face images from datasets like MS-Celeb-1M [9], Glint-360k [10], and IJB-C [11], thereby missing the interdependence between the face detection and recognition modules.

Several approaches [12, 13, 14] also propose end-to-end face detection and recognition systems. However, in all these cases the detection and recognition modules are evaluated either on extremely simple face recognition datasets that do not have challenging detection scenarios or on publicly available datasets for each of the individual tasks. Both these cases can lead to misinterpretation of system performance estimates. This also shows

---
This work has been supported by the Ford Motor Company and SAIPS under the Ford-Purdue Alliance Program.

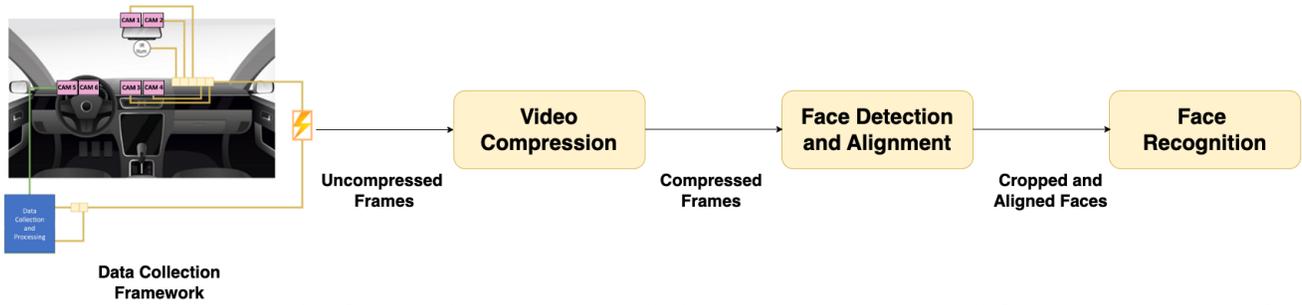

Figure 1: An end-to-end practical face analytics system deployed inside a vehicle. Collected data is first compressed and then, face detection, alignment, and recognition are performed sequentially.

the lack of suitable datasets for end-to-end evaluation of practical video analytics systems.

In this paper, we show that it is important to use balanced datasets with consistent annotations across several interdependent tasks of a practical video analytics system to obtain meaningful and interpretable performance estimates.

## End-to-end Face Analytics

Our practical face analytics system shown in Figure 1 performs face detection, alignment and recognition sequentially in a driving-specific scenario. The data collection framework of our system captures occupants inside a vehicle under different illumination conditions using both RGB and IR cameras mounted at different locations inside the vehicle. The collected data is then compressed using popular video codecs like AVC [15] and HEVC [16]. Next, the compressed data is passed onto the detection, alignment, and recognition modules of our face analytics system in a sequential order. During our end-to-end evaluation, we ensure that each module of our system operates only on the outputs of the previous modules in order to capture task interdependence.

In the next few sections, we will look into the details of each block of our face analytics system and the steps required at each stage to perform a meaningful end-to-end evaluation.

### *Data Collection Framework*

To collect suitable driving-specific data for the evaluation of our system, we use the framework in Figure 2. This setup uses cameras of different modalities, i.e., RGB and IR, mounted at three different locations inside the vehicle: the console, the rear-view mirror and the wheel. It is used to capture 50 vehicle occupants of various ages, ethnicities, and skin tones, inside a vehicle in different outdoor (well-lit) and indoor (low-light) illumination conditions. To use this data for the evaluation of face detection and recognition, it is annotated by humans in terms of face identities, bounding boxes, facial landmarks, and pose angles. Sample frames captured using this framework are shown in Figure 3.

Figure 4 shows the characteristics of our dataset in terms of the different scenarios that we consider for our end-to-end evaluation, as well as face pose characteristics like yaw, pitch and roll angles. For our evaluation, we estimate our system performance by grouping different scenarios in the dataset as follows :

- Indoor vs Outdoor illumination conditions;
- RGB vs IR camera modalities;
- Console vs Rear-View Mirror vs Wheel camera locations.

Figure 4 also illustrates that the dataset has an extremely large number of samples which greatly increases the time complexity of performing an end-to-end evaluation. For example, to perform an exhaustive face recognition evaluation, $10^{12}$ face pairs need to be considered. Furthermore, the samples in each evaluation group are imbalanced which can lead to an unwanted bias in performance estimates.

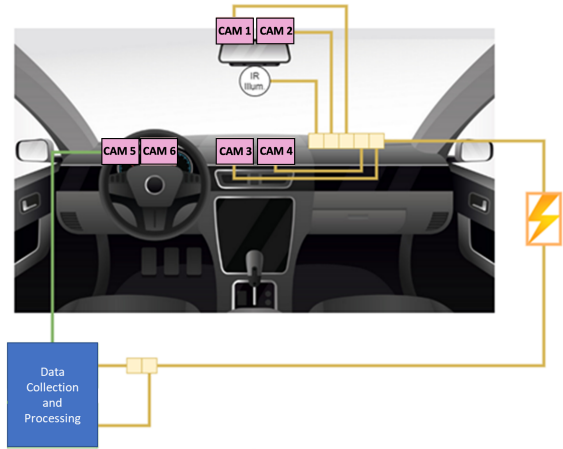

Figure 2: Multi-camera, multi-modal data collection framework use to capture faces of vehicle occupants.

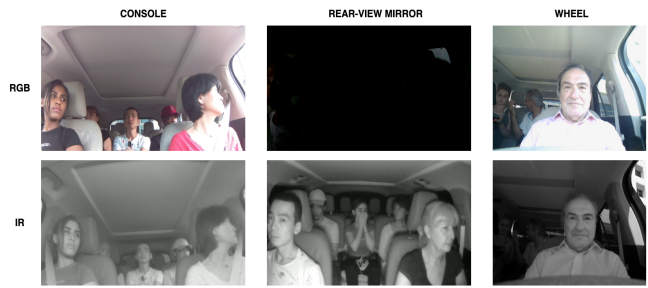

Figure 3: Occupants captured inside a vehicle by the data collection setup in Figure 2. From left to right, we see samples from the console, rear-view and wheel camera. The top and bottom rows show the RGB and IR modalities of the same scene respectively and there exists a slight offset between the modalities. The console/wheel samples and rear-view samples show the outdoor/well-lit and indoor/low-light illumination scenarios, respectively.

Another challenge that we encounter with our dataset, is that its annotations are inconsistent across different camera modalities and lighting conditions. This is mainly because 1) there exists a slight offset between the RGB and IR cameras at each camera location (Figure 3), and 2) the human generated annotations fail to capture the actual number of occupants inside the vehicle in certain camera modalities and lighting conditions. For example, in Figure 5, the RGB stream in the indoor low-light scenario has a single face annotation when in actuality there are four occupants inside the vehicle as seen in the IR stream. When such inconsistent annotations are used as ground-truth during evaluation, they create misleading system performance estimates.

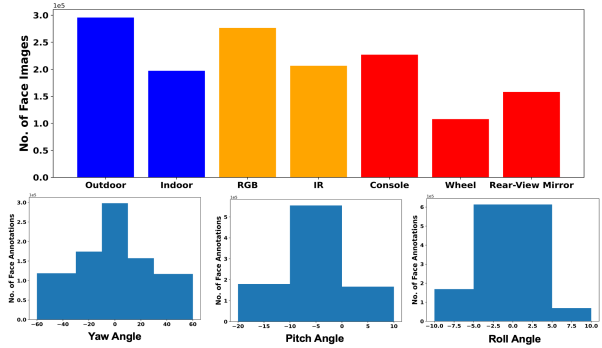

Figure 4: Distribution of samples in our original driving-specific dataset; Large number of samples with imbalances in the evaluation scenarios.

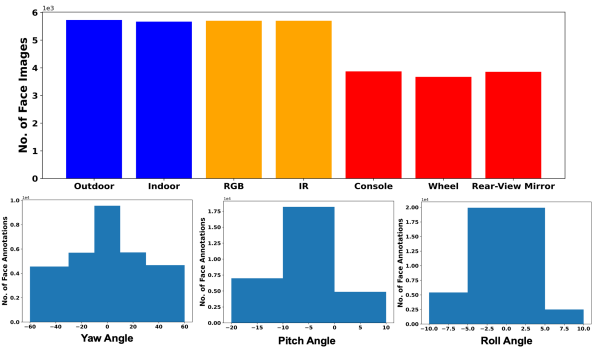

Figure 6: Distribution of samples in a subset created using our balancing strategy; Face pose characteristics are retained while the number of evaluation samples are reduced and balanced across evaluation pairs.

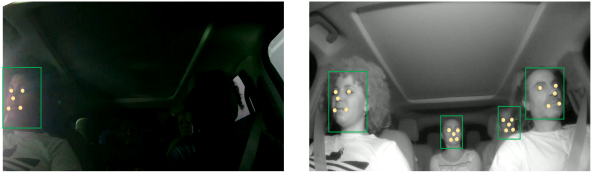

Figure 5: Inconsistent human annotations across camera modalities and illumination conditions lead to incorrect end-to-end performance estimates.

Here, we propose several procedures that create smaller balanced subsets of our dataset while retaining important characteristics of the original dataset like face pose angles. These also fix human-generated annotations such that they are consistent across different camera locations, camera modalities, and illumination conditions to enable an accurate evaluation of our face analytics system.

*Dataset Pre-processing*

To perform an end-to-end evaluation that is fast and accurate, we create smaller balanced subsets of our dataset using a simple and effective iterative balancing strategy. The samples of these subsets are balanced across all the scenarios in each evaluation group, to ensure any imbalances are eliminated. Furthermore, our strategy ensures that the subsets retain the original dataset's face pose characteristics in terms of yaw, pitch, and roll angles. These characteristics have a significant impact on the system performance estimates, and altering these can lead to performance estimates that do not necessarily reflect the performance of the system on real-world data. We demonstrate this later in the *Experiments and Results* section.

Our balancing strategy considers one face identity at a time and populates bins for each evaluation scenario while ensuring that an equal number of samples are chosen for each evaluation scenario. Furthermore, the samples selected for each face identity have similar face pose statistics as the original dataset. Operating on a single face identity at a time allows our strategy to create a subset that has an equal number of samples for all face identities in all the evaluation scenarios for both face detection and recognition, while also retaining the diverse face pose characteristics of the original dataset.

Figure 6 shows the characteristics of a subset created using our strategy. This subset has fewer evaluation samples, does not have imbalances in terms of samples per evaluation group, and also has similar face pose characteristics as the original dataset. In the *Experiments and Results* section, we show that we obtain similar performance estimates of our system over several non-overlapping subsets created using this strategy, as long as the face pose characteristics are retained.

*Fixing Dataset Annotations*

As humans annotate our dataset, we face additional challenges while evaluating our system. Often, the annotations are inconsistent across lighting conditions and camera modalities. They are also misaligned due to the offsets that exist at each camera location. In order to perform a meaningful evaluation of our face analytics system, we must fix these inconsistent annotations.

We see the offset between the camera modalities at each location and inconsistencies in human-generated face annotations in Figures 3 and 5 respectively. These annotations lead to incorrect system performance estimates, which can be dangerous when the system is being deployed in the real-world. Hence, it is necessary to correct these issues and create a single set of consistent annotations that represent the exact number of faces inside the vehicle at any given time, irrespective of the camera location, modality, and lighting condition.

To fix the annotations, we first eliminate the offset between the RGB and IR cameras using a popular transformer-based feature-matching algorithm, LoFTR [17], that finds keypoints between the same scene captured in the different camera modalities. These keypoints are used to estimate a homography matrix to align the RGB and IR face annotations. Next, we eliminate redundant overlapping annotations to create a single set of consistent annotations that captures the exact number of faces in any scene, irrespective of the camera modality and illumination condition. Figure 7 shows an example of how inconsistent annotations are fixed using our procedure for a given scene across different camera modalities and lighting conditions.

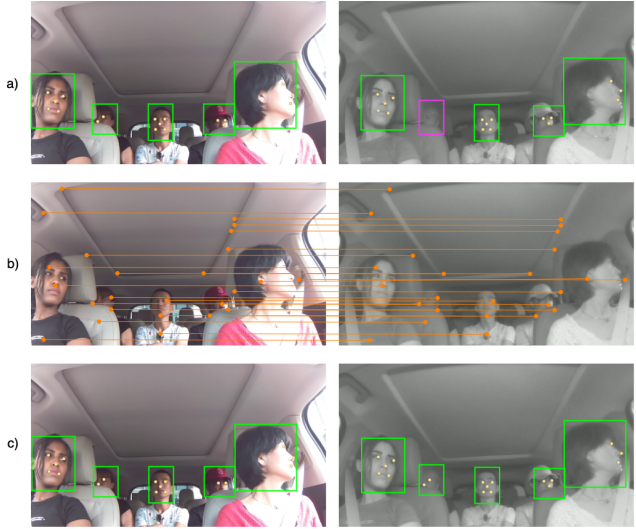

Figure 7: Fixing annotations across different camera modalities and illumination conditions; **a)** shows the offset between the camera modalities and a missing face annotation in the IR stream. **b)** shows the LoFTR keypoints used to estimate a homography to align the annotations. **c)** shows a consistent set of annotations suitable for both the camera modalities and lighting conditions.

With a smaller balanced subset of our dataset which 1) retains the original dataset's face pose characteristics and 2) has consistent annotations across different camera locations, modalities and lighting conditions, we can now evaluate the different face analytics tasks of our system in an end-to-end manner to obtain accurate performance estimates.

### *Video Compression*

In practical video analytics, compression is a necessity in order to meet bandwidth constraints that exist in real-world scenarios. Popular video codecs like AVC [15] and HEVC [16] are used extensively in such cases in order to meet these constraints. Most video analytics today rely on deep learning models that are usually trained on high quality inputs. These models are robust to small amounts of compression, but as the level of compression increases, the performance of the analytics decreases rapidly [18]. Thus, in our experiments, we demonstrate how video compression affects the overall performance of face analytics by including different levels of compression in our evaluation.

### *Face Detection and Recognition*

Our practical face analytics system performs face detection, alignment and recognition. In our system, we ensure that each of these modules are evaluated on the output of the other in a sequential manner to capture task interdependence.

For face detection and alignment, we use the RetinaFace [2] detector with the ResNet50 [19] backbone. We use a pre-trained model trained on the WiderFace [4] dataset and evaluate it on our balanced driving-specific data subset. Our detection model operates on compressed data inputs and predicts face bounding boxes and landmarks. These outputs are used to crop and align the face images using similarity transforms before they are passed onto our recognition module.

For face recognition, we rely on the LResNet-100E-IR [6] model trained using the ArcFace [6] loss. Similar to the detection case, we rely on pre-trained models that have been trained on several large face recognition datasets like Glintk360k [10] and MS-Celeb-1M [9]. During evaluation, instead of following the typical approach in which perfect detection is assumed, we use the outputs of our detection module. This ensures that the detection performance impacts recognition performance, and specifically, that missed detections negatively impact recognition performance.

In real-world scenarios like driving analytics inside a vehicle, there are also resource constraints that must be considered. Using complex deep learning models for analtyics might not always be the best solution. Therefore, we also estimate the end-to-end performance of our face analytics system by replacing its complex learning models with lightweight and efficient MobileNet ([20], [21]) counterparts, to determine if such lightweight models can be effectively used for practical face analytics.

## Experiments and Results

In this section, we demonstrate that 1) using balanced subsets of our driving-specific dataset with similar face pose characteristics, 2) creating a single set of consistent annotations suitable for all camera modalities and lighting conditions, and 3) capturing task interdependence by evaluating face analytics tasks sequentially all improve the interpretability and accuracy of the system performance estimates. We also evaluate the effects of compression on the performance of our system and the effectiveness of lightweight deep-learning models in practical video analytics.

As seen in the *Data Collection Framework* section, we group the different scenarios of our driving-specific dataset based on illumination condition, camera modality, and camera locations to help make our evaluation results more interpretable. We encode our dataset using x265 [16] at different compression levels by picking six quantization parameters (QP) in the range of 18-50 during encoding. These QPs control the extent of compression applied to the dataset; the larger the QP, the more the compression. In all our experiments, we report our detection model's performance in terms of mean-Average-Precision (mAP) [22] and for face recognition, we perform 1:1 verification [11] and report the True Positive Rate (TPR) at a False Positive Rate (FPR) of 0.01.

First, we look at how creating smaller balanced subsets of our dataset with similar face pose characteristics as the original dataset helps us to obtain accurate end-to-end performance estimates. Using our balancing strategy from the *Dataset Preprocessing* section, we create multiple non-overlapping subsets with characteristics similar to those in Figure 6. We then estimate the detection performance on these subsets and their compressed versions. Figure 8 shows us the results obtained on a single subset. We see a significant drop in the detection performance in all scenarios as the bandwidth decreases and the compression increases. We obtain similar performance estimates on multiple non-overlapping subsets as long as our these subsets have the same face pose characteristics as the original dataset.

To show how face pose characteristics can affect system performance estimates, we evaluate the detector performance on a balanced subset whose face pose characteristics have an equal number of samples in each of the yaw, pitch, and roll bins in-stead of the original dataset's characteristics. This clearly does not reflect the actual distribution of face poses encountered in the dataset. Figure 9(a) shows us the detection results obtained on this subset. It is clear from this figure that the detection performance is underestimated in almost all scenarios. This is because harder face samples are now used for evaluation instead of the actual face poses typically encountered in practice. Thus, in order to obtain more accurate performance estimates from smaller balanced subsets, it is essential to retain the original face pose characteristics.

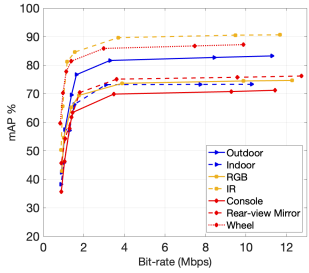

Figure 8: Detection mAP% over balanced subsets of the driving-specific dataset with original face pose characteristics.

Similarly, we demonstrate why consistent dataset annotations are necessary to perform an accurate end-to-end evaluation; that is, the annotations should be consistent irrespective of the camera location, modality or lighting condition. In this case, we estimate the performance of the detection module on a balanced subset of our dataset while using two different sets of annotations as ground-truth during evaluation; one with human-generated annotations, and the other with annotations that are consistent. The results of this experiment are shown in Figure 9(b). In this case, we look at the results for the Indoor and RGB evaluation scenarios as these have the highest number of inconsistent annotations (Figure 5). From our results, we see that using human-generated annotations as ground-truth during evaluation overestimates the system performance. This is because these annotations significantly reduce the number of faces considered during evaluation. Such misleading performance estimates can be extremely dangerous when the system is deployed in the real-world. When we fix our annotations, we obtain a more accurate and realistic performance estimate of our detection module.

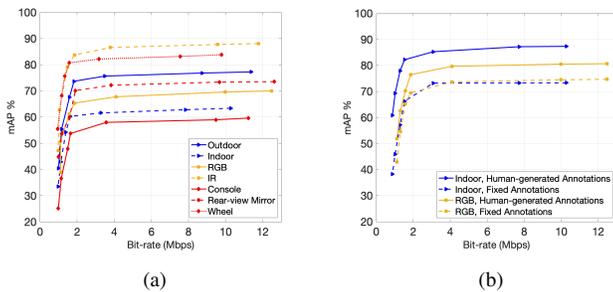

(a)　　　　　　　　　　(b)

Figure 9: Misleading detection performance estimates due to a) altered face pose characteristics and b) inconsistent annotations.

Next, we demonstrate task interdependence must be taken into account while estimating end-to-end system performance. Recognition models are usually evaluated under the assumption that all the faces in an image are perfectly detected and aligned. However, in an end-to-end practical system, it is dangerous to assume perfect face detection. As seen earlier in Figure 8, the detection performance is not perfect and varies based on the evaluation scenario. To demonstrate the interdependence between detection and recognition, we evaluate the performance of our recognition module on two different sets of inputs: 1) assuming perfect detection for all faces in the dataset, and 2) on the outputs of the detection module. The inputs are aligned using similarity transforms before recognition. Figure 10 illustrates that when we assume perfect detection (Figure 10(a)), the recognition module performance is significantly overestimated in all scenarios. However, when we consider the detection module's outputs during evaluation (Figure 10(b)), we get a truer sense of the recognition module's performance. The results, in this case, are in agreement with the detection results (Figure 8); specifically, the recognition performance estimates depend on how well the detection module performs in every scenario. In contrast, when we consider perfect detection in Figure 10(a), we see that the recognition module performs extremely well in certain scenarios (for example, Console Camera) even though the actual detection performance in these scenarios is poor. Thus, capturing task interdependence while evaluating practical end-to-end video analytics systems is critical for accurate performance evaluation.

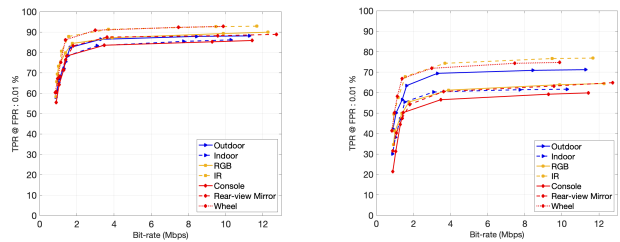

(a) Assuming perfect detection　　(b) Using detection outputs

Figure 10: Recognition performance varies significantly based on the inputs used during evaluation.

Finally, because video analytics systems are deployed in resource constraint environments like autonomous vehicles, we also evaluate the end-to-end performance of our analytics system when it uses lightweight deep-learning models for face detection and recognition. To reduce the model parameters, we replace the backbone of the RetinaFace detector with MobileNet0.25 [20]. Similarly, we replace the LResNet-100E-IR [6] with the LMobileNetE [21] model for recognition. The end-to-end performance estimates using these lightweight models are seen in Figure 11. From these plots, we infer that in comparison to the results seen in Figures 8 and 10(b) respectively, the lightweight models see an overall performance drop. Interestingly the relative end-to-end performance estimates stay consistent across all scenarios.

## Conclusion

In this paper, we demonstrate the importance of a careful performance analysis of practical analytics systems. This requires (a) end-to-end assessment (b) using balanced dataset subsets (c) with consistent annotations. We propose several strategies to perform a this evaluation. We consider how well they behave in practical bandwidth and resource constraint environments by evaluating their robustness to video compression and replacing the complex models in these systems with their lightweight counterparts, respectively. Through our experiments, we obtain accurate and interpretable performance estimates that can be effective in improving system performance in practical scenarios.

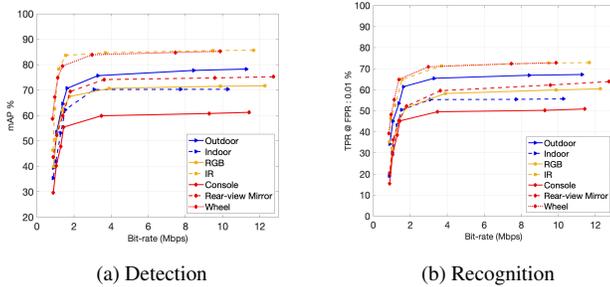

(a) Detection      (b) Recognition

Figure 11: System performance using lightweight models.